\documentclass[letterpaper, 10 pt, conference]{ieeeconf}  % Comment this line out
\IEEEoverridecommandlockouts                              % This command is only
\usepackage{graphicx}

\title{\large\textbf{
ADVANCING RECOMMENDER SYSTEMS BY MITIGATING SHILLING ATTACKS
}}

\author{Aditya Chichani$^{*}$, Juzer Golwala$^{*}$, Tejas Gundecha$^{*}$, Kiran Gawande$^{*}$
\\Computer Engineering Department, Sardar Patel Institute of Technology
\thanks{*All authors have contributed equally to this work}% <-this % stops a space
}
\begin{document}
\onecolumn{
\textsuperscript{\textcopyright} 2018 IEEE. Personal use of this material is permitted. Permission from IEEE must be obtained for all other uses, in any current or future media, including reprinting/republishing this material for advertising or promotional purposes, creating new collective works, for resale or redistribution to servers or lists, or reuse of any copyrighted component of this work in other works.
\newline
\newline
This paper was accepted to be published in IEEE, Proceedings of 2018 9th International Conference on Computing, Communication and Networking Technologies (ICCCNT) 
\newline
\newline 
DOI 10.1109/ICCCNT.2018.8494141
}
\twocolumn{}
\maketitle
\thispagestyle{empty}
\pagestyle{empty}

%%%%%%%%%%%%%%%%%%%%%%%%%%%%%%%%%%%%%%%%%%%%%%%%%%%%%%%%%%%%%%%%%%%%%%%%%%%%%%%%
\begin{abstract}
Considering the premise that the number of products offered grow in an exponential fashion and the amount of data that a user can assimilate before making a decision is relatively small, recommender systems help in categorizing content according to user preferences. Collaborative filtering is a widely used method for computing recommendations due to its good performance. But, this method makes the system vulnerable to attacks which try to bias the recommendations. These attacks, known as \lq{shilling attacks}' are performed to push an item or nuke an item in the system. This paper proposes an algorithm to detect such shilling profiles in the system accurately and also study the effects of such profiles on the recommendations.

\end{abstract}
%%%%%%%%%%%%%%%%%%%%%%%%%%%%%%%%%%%%%%%%%%%%%%%%%%%%%%%%%%%%%%%%%%%%%%%%%%%%%%%%
\section{INTRODUCTION}

Recommendation Systems have helped users in filtering important data from the large pool of data available on the internet. It has also helped organizations in targeting appropriate users for selling their items. All this has been made possible due to various statistical and machine learning models which try to predict what rating a user would give an item which he has not come across. Then on the basis of this ‘rating’ or ‘preference’ items can be recommended to a user. Hence Recommender Systems have become a powerful business strategy which helps the end-user in deciding what to buy and for the organizations to pitch their products to the appropriate users.

\section{BACKGROUND}

\subsection{Collaborative Filtering(CF)}

Collaborative filtering techniques[3] predict the rating of an item by computing similarity between users or items (i.e.by forming collaboration of users or items). Collaborative filtering is carried in two ways:

\subsubsection{User-User CF}
In user-user CF similarity between two users is computed using the user-item ratings matrix. Similarity between \lq{user1}' and \lq{user2}' is computed using Cosine-Distance wherein \lq{n}' items are taken into consideration which have been rated by both the users. Then once the similarity is computed, we can predict rating of item \lq{user1}' which has been rated by \lq{user2}' based on the similarity computed. We do this for all the users in the database and then predict rating of a specific item \lq{m}' for user \lq{k}' using the following formula: 

\begin{figure}[ht]
\centering
\includegraphics[width=3 in]{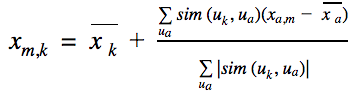}
\end{figure}

where the first term is the mean rating of user \lq{k}’ which is used for normalization. It defines how a particular user rates an item. There maybe users who give a rating 3 to items they dislike and some users would give a rating of 1(on a scale of 1 to 5). Hence this term is essential for predicting rating for each different user. The denominator in the second term is again used for normalization so that the predicted rating lie in the range of 1 to 5.[3]

\begin{figure}[ht]
\centering
\includegraphics[width=3.6in]{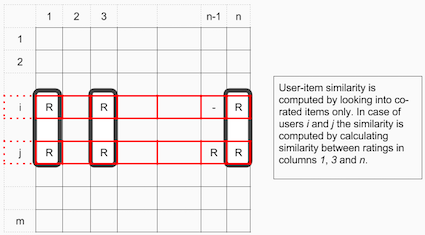}
\caption{Similarity between users}
\label{figurelabel}
\end{figure}

\subsubsection{Item-Item CF}
Item-Item CF computes similarity[3] between two items. Similarity between two items, \lq{item1}' and \lq{item2}' is computed by considering \lq{n}' users who have rated both the items and then computing Cosine-Distance. In this case if a user likes \lq{item1}' and has not come across \lq{item2}', and \lq{item1}' and \lq{item2}' turn out to be similar, \lq{item2}' can be recommended to that particular user. The rating for item \lq{m}' by user \lq{k}' can be predicted using the following formula: 
\begin{figure}[ht]
\centering
\includegraphics[width=2in]{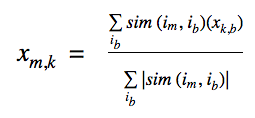}
\end{figure}

In this formula, similarity of item \lq{m}' is computed with each item \lq{b}' in the database, then depending on the rating given to item \lq{b}' by user \lq{k}', rating for item \lq{m}' is computed. The denominator is used as a normalizing factor so that the predicted ratings lie in the range of 1 to 5.
\begin{figure}[ht]
\includegraphics[width=3.4 in]{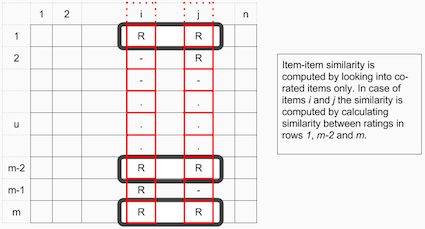}
\caption{Similarity between items}
\label{figurelabel}
\end{figure}
\pagebreak
\subsection{Content Based Filtering}

Content based filtering model recommends each individual user according to his \lq{taste}' by going through the items previously liked by the user. It creates a separate profile for each user on the basis of the previously rated items’ genre. For example, if a particular user has highly rated a lot of horror movies, the user will be probably be recommended \lq{The Conjuring}' as well. Content based recommender systems usually ask the user to fill a survey to learn his/her tastes which eliminates the \lq{Cold Start}' problem.
 In a Collaborative Filtering recommender system,when a new user enters the system,similarity cannot be computed due to lack of any entries in the rating matrix for that particular user and as a result the recommendations for that user will be vague. This problem is commonly referred to as the \lq{Cold Start}' problem.
However, Content Based filtering systems suffer from over-specialization since a user who has previously watched and rated thriller movies highly may end up getting recommended only thriller movies although his taste might not be restricted to only thriller movies
In practice, CF systems have better performance than content based systems and are more widely used. Another category of recommender systems is the Model Based recommender systems which are hybrid in nature.[3]

\begin{figure}[ht]
\centering
\includegraphics[width=3.3in]{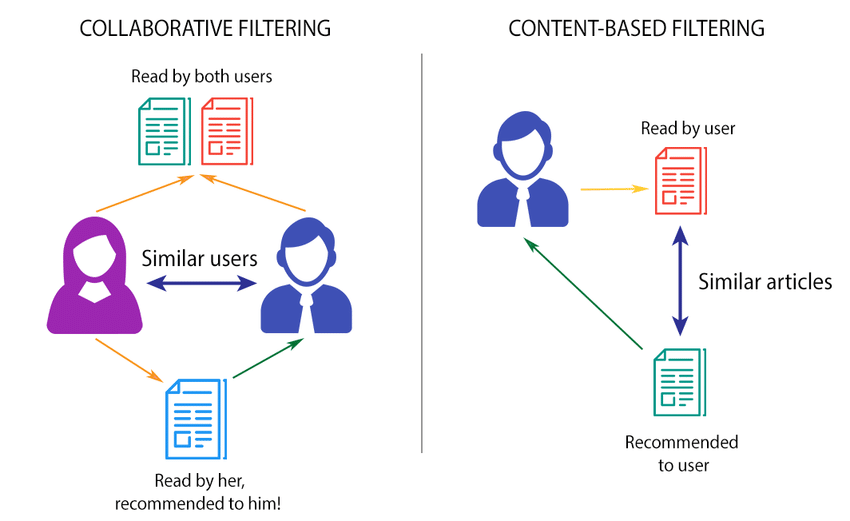}
\caption{Difference between collaborative and content based filtering}
\label{figurelabel}
\end{figure}

\subsection{Model Based Filtering}
Model based filtering [3] makes use of the user ratings matrix to find latent or hidden features of the items that a particular user tends to favour just like in content based filtering and also uses the similarities  between users to avoid overspecialization. To achieve this, Model based filtering uses a matrix factorization technique like SVD (Singular Value Decomposition).
\newline
The equation for SVD can be represented as follows:
\newline
X = USV$^{T}$ 
\newline
where for a given m x n matrix X:
\begin{itemize}
\item U is an (m x r) orthogonal matrix
\item S is an (r x r) diagonal matrix with non-negative real numbers on the diagonal
\item V$^{T}$ is an (r x n) orthogonal matrix
\end{itemize}

Elements on the diagonal in S are known as singular values of X.
Matrix X can be factorized to U, S and V.
 The U matrix has the users as rows and the hidden feature vectors as columns while the V matrix has the hidden feature vectors as rows and the items as columns. These hidden feature vectors are actually eigen vectors and the singular values in X are the eigen values for the corresponding eigen vectors.

The predictions in Model based filtering are obtained by computing the dot product of U,S and V$^{T}$ 

\begin{figure}[ht]
\centering
\includegraphics[width=3in]{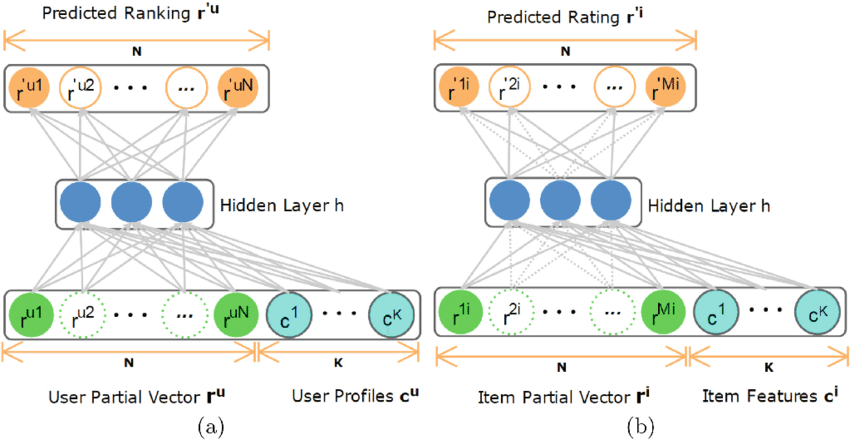}
\caption{Model Based Filtering}
\label{figurelabel}
\end{figure}

\subsection{Shilling Attack}
Recommendations are computed by calculating similarity on the basis of users or items. If the recommender system takes into consideration similarity between users while computing recommendations, there may be a few profiles which are not genuine and try to bias the recommender system. These profiles are shilling profiles. These profiles may aim to promote their item by rating it highly or demote item of their competitor.[3,5]

The structure of user profiles in recommender systems are such that it makes the system vulnerable to such attacks and the attacker profiles go undetected. These profiles bias the recommendations and the users may not like the recommendations anymore in turn leading to financial losses. The general structure of a shilling profile is shown in Fig.5 

\begin{figure}[ht]
\centering
\includegraphics[width=3.3in]{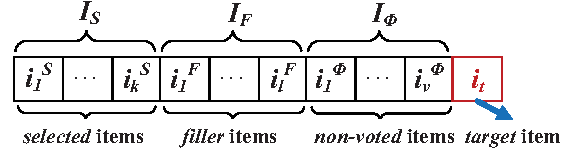}
\caption{The illustration of a shilling profile}
\label{figurelabel}
\end{figure}

% \begin{figure*}[ht]
% \centering
% \includegraphics[width=\textwidth]{image10.png}
% \caption{The illustration of a shilling profile}
% \label{figurelabel}
% \end{figure*}

The list of selected items contains those items which a profile uses to get into the neighbourhood of large chunks of users. These are the popular items which are rated positively by many users. List of filler items are the items which a profile would use to mask itself as genuine user. List of target items are the items which an attacker is actually interested in. The attacker would either rate the target item highly if he wants to push the item or rate it poorly if he wants to nuke the item. These are the items whose RMSE values are affected, i.e the difference in prediction of rating for those items before and after the attack is very high.[6]

 These attacks are performed using various models such as:

\subsubsection{Random Attack Model}
In Random Attack, the group of attacker profiles select the filler items randomly from the rating database and give these items a mean rating of the entire database. The target items are given same rating by all the profiles, i.e high rating for push attack and poor rating for nuke attack.[5,6]

\subsubsection{Average Attack Model}
In Average Attack, the group of attacker profiles select the filler items randomly from the rating database and give these items a mean rating of that particular item. The target items are rated in a similar manner to random attack.[5,6]

\subsubsection{Bandwagon Attack Model}
In Bandwagon Attack, filler items are selected randomly and can be rated as in Random Attack or Average Attack. The target items are also rated similar to previous attacks. The main difference in this attack is the use of an additional list of selected items. These items are the ones which are rated by many users in the system, i.e they are popular and also the ratings are positive. Hence an attacker group would rate these items highly to get into the neighbourhood of large chunks of users. This is a very powerful attack in terms of recommendations which are calculated on the basis of similarity between users.[5,6]

\subsubsection{Segment Attack Model}
Segment attacks require additional knowledge about items as to which categories they belong to. In this attack, the group of attackers would rate items similar to target items highly so that they can get into the cluster of users who like  items of that category and can target those users. Hence, this attack is performed on a segment of users who like items of the same category as that of target item.
\linebreak
\linebreak
These attacks tend to give false recommendations and decrease the trust of existing users of the system. Hence, proper detection algorithms are necessary to detect profiles trying to perform shilling attack.

\section{Proposed Detection Algorithm}
In this section, the proposed method is discussed.We are using unsupervised learning approach for shilling attack detection so that this detection method can be applied to any general recommender system. Based on the concept that shilling profiles have high covariance with other profiles , PCA is applied to detect shilling profiles.
\newline\newline
Algorithm:

\begin{figure}[ht]
\centering
\includegraphics[width=3.7 in]{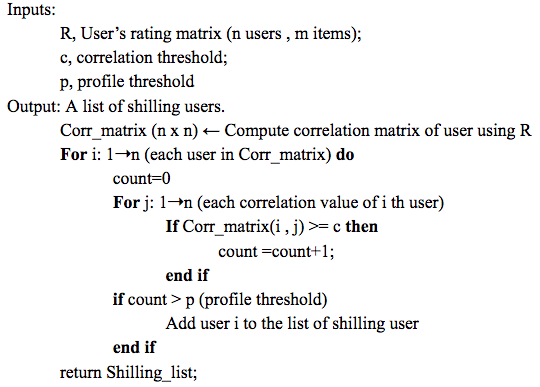}
\caption{PCA Algorithm}
\label{figurelabel}
\end{figure}

Value of Correlation threshold should be very high as shilling profiles have very high correlation with other users. Setting low value for correlation threshold might detect genuine profiles as shilling profiles which decreases accuracy of detection. Profile threshold is the number of users which can affect the recommendations and make system biased. Some heuristic methods  are able to find profile threshold.

\begin{figure}[ht]
\centering
\includegraphics[width=3.5in]{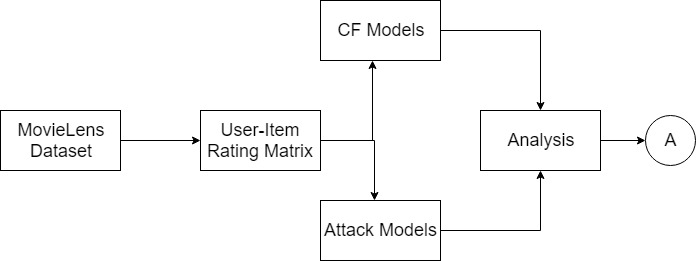}
\end{figure}

\begin{figure}[ht]
\centering
\includegraphics[width=3.5in]{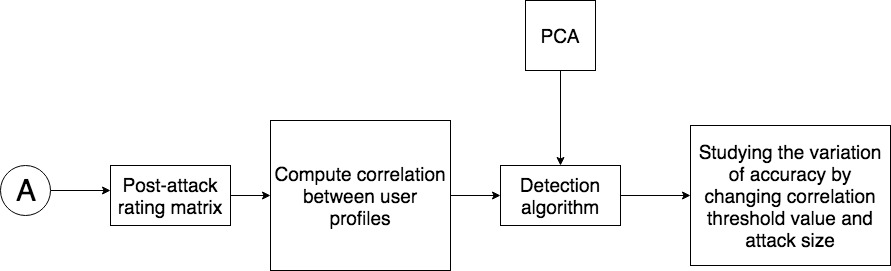}
\caption{Block Diagram}
\label{figurelabel}
\end{figure}

\par
In this proposed method, PCA detection is applied to detect shilling profiles. Initially correlation is computed between the users from Rating matrix. Correlation matrix has these correlation values of each user with all other users. In PCA, profiles having high correlation with many other profiles is considered as shilling. If profile is having correlation greater than correlation threshold with other profiles and is having count greater than profile threshold,then that profile is detected as shilling profile.

\par
Shilling profiles tend to have high correlations as this attack is performed in groups. Hence, these profiles have rated the target items similarly and also the filler items which are selected randomly have a similar rating. If an attacker group would try to rate the filler items differently, i.e different ratings for same filler items, it would increase the cost of the attack. Hence, the assumption that shilling profiles have a high correlation gives more or less accurate result if the profile threshold is set correctly.
\section{Experimental Discussion}
In this section, accuracy of the proposed algorithm is evaluated. MovieLens 100K dataset [5,6] containing 943 user profiles and 1682 movie items is used in the experiment. This dataset contains overall 100K ratings ranging from 1 to 5. Rating 5 is the maximum allowed rating while rating 1 is minimum allowed rating.
	Shilling profiles have been generated using different attack models like random , average and bandwagon attack with different values of attack size (5\%  to 25\%) and filler size( 5\% , 10\% , 15\%).Based on above shilling profiles , a new dataset containing authentic profiles and shilling profiles is generated. The parameter correlation threshold and profile threshold is assumed to be 0.95 and 10\% of the authentic users respectively.
	The performance of the detection method is calculated using a parameter known as F-measure[5,6].It is a measure of test's accuracy. F-measure is harmonic mean of precision and recall.
\newline
\begin{figure}[ht]
\centering
\includegraphics[width=3 in]{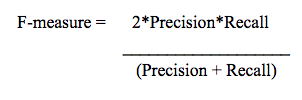}
\end{figure}

Precision is defined as the ratio of number of detected true shilling profiles to the total number of shilling profiles returned by the detection algorithm where Recall is the ratio of the detected true shilling profiles to the total number of actual shilling profiles present in the system.
\linebreak

\begin{figure}[ht]
\centering
\includegraphics[width=3in]{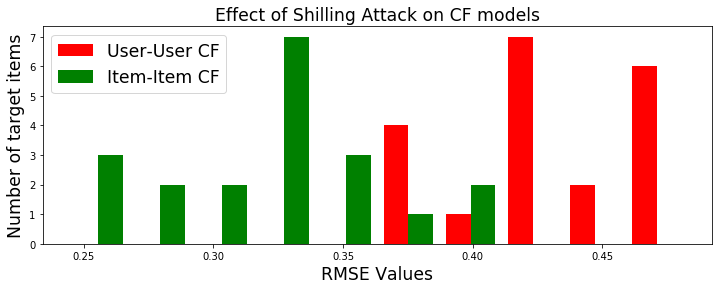}
\caption{Effect of shilling attack on RMSE of target items}
\label{figurelabel}
\end{figure}
\pagebreak
\begin{figure}[ht]
\centering
\includegraphics[width=2.5in]{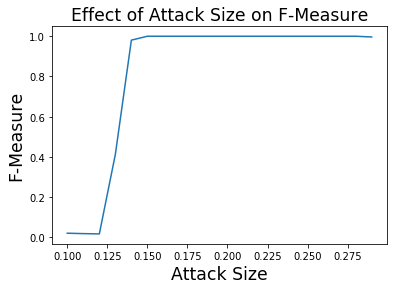}
\caption{F-Measure for Random Attack}
\label{figurelabel}
\end{figure}

\begin{figure}[ht]
\centering
\includegraphics[width=2.5in]{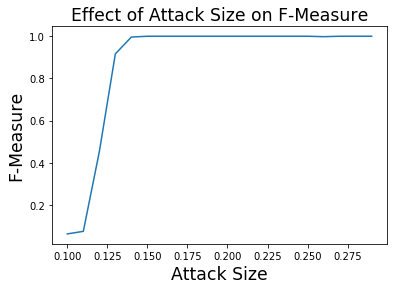}
\caption{F-Measure for Average Attack}
\end{figure}

\begin{figure}[ht]
\centering
\includegraphics[width=2.5in]{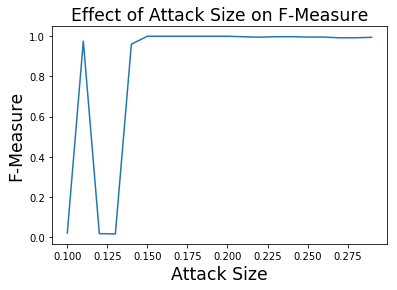}
\caption{F-Measure for Bandwagon Attack}
\label{figurelabel}
\end{figure}

\section{CONCLUSIONS}
On calculating the RMSE values for target items, it was found that when the recommendations are computed on basis of user similarity, they are affected more than the recommendations that were computed on basis of item similarity. Hence, while implementing a recommender system which considers similarity between items, shilling attack detection may not be performed frequently. But, recommendations computed on basis of user similarity have better accuracy and then shilling attack detection should be taken care of. From the experimental results, we can deduce that when the number of shilling profiles in the system are greater than or equal to profile threshold, detection algorithm will run with a very high accuracy. Conversely, if the shilling profiles are less than profile threshold, recommender system will not be affected.

\section{Relation to Prior Work}
This paper aims to research and establish the effect of different shilling attack models on various recommendation techniques.The work documented by Xiaoyuan Su and Taghi M. Khoshgoftaar [3] compares the recommendation models in the scenario where no attacks have been performed.While the technique proposed by Zi-Jun Deng et al [5] can be used to mitigate shilling attacks,not much information is given on how the accuracy of different models is affected due to it. 

\section{Motivation}
Unsupervised learning algorithms with a high accuracy are better than supervised learning algorithms since it is difficult to gather data with balanced classes for shilling attacks. Also for each recommendation engine, the structure of user profile would change and with it the training data would change. Hence, to mitigate this attack we need an unsupervised approach so that it can work with any recommendation model. Hence, the PCA approach mentioned in this paper gives us a robust unsupervised approach to detect shilling profiles in the system.

\addtolength{\textheight}{-12cm}   % This command serves to balance the column lengths
                                  % on the last page of the document manually. It shortens
                                  % the textheight of the last page by a suitable amount.
                                  % This command does not take effect until the next page
                                  % so it should come on the page before the last. Make
                                  % sure that you do not shorten the textheight too much.

%%%%%%%%%%%%%%%%%%%%%%%%%%%%%%%%%%%%%%%%%%%%%%%%%%%%%%%%%%%%%%%%%%%%%%%%%%%%%%%%

%%%%%%%%%%%%%%%%%%%%%%%%%%%%%%%%%%%%%%%%%%%%%%%%%%%%%%%%%%%%%%%%%%%%%%%%%%%%%%%%

%%%%%%%%%%%%%%%%%%%%%%%%%%%%%%%%%%%%%%%%%%%%%%%%%%%%%%%%%%%%%%%%%%%%%%%%%%%%%%%%
\section*{APPENDIX}
\subsection{RMSE}
RMSE stands for Root Mean Squared Error 
which is computed between predictions(recommendations) before and after the shilling attack.

\subsection{Push Attack}
These are shilling attacks wherein a group/user tries to promote their own item in the system by giving the item maximum allowed rating.

\subsection{Nuke Attack}
These are shilling attacks wherein a group/user tries to demote item of an adversary by giving the item minimum allowed rating.

%%%%%%%%%%%%%%%%%%%%%%%%%%%%%%%%%%%%%%%%%%%%%%%%%%%%%%%%%%%%%%%%%%%%%%%%%%%%%%%%


\begin{thebibliography}{99}
\bibitem{c1}Zhihai Yang and Zhongmin Cai `Detecting abnormal profiles in collaborative filtering recommender systems', \textit{J. Intell. Inf. Syst.},\hskip 1em plus 0.5em minus 0.4em\relax Yang C17, India, pp. 499--518,2017.
\bibitem{c2} Gunes, Ihsan and Kaleli, Cihan and Bilge, Alper and Polat, Huseyin, `Shilling Attacks Against Recommender Systems: A Comprehensive Survey,' \emph{Artif. Intell. Rev.},\hskip 1em plus 0.5em minus 0.4em\relax vol. 42, pp. 767--799, Dec 2014.
\bibitem{c3} Xiaoyuan Su and Taghi M. Khoshgoftaar, `A Survey of Collaborative Filtering Techniques' \emph{Hindawi Publishing Corporation Advances in Artificial Intelligence },\hskip 1em plus 0.5em minus 0.4em\relax Volume 2009, Article ID 421425, 19 pages, August 2009.
\bibitem{c4}Wei Zhou , Junhao Wen , Qingyu Xiong , Min Gao, Jun Zeng, `SVM-TIA a shilling attack detection method based on SVM and target item analysis in recommender systems,'\textit{Elsevier}, 19 October 2016 
\bibitem{c5} Zi-Jun Deng , Fei Zhang , Sandra P. S. Wang , `Shilling Attack Detection In Collaborative Filtering Recommender System By PCA Detection And Perturbation', \textit{2016 International Conference on Wavelet Analysis and Pattern Recognition (ICWAPR)},\hskip 1em plus 0.5em minus 0.4em\relax pp.213-218, July 2016
\bibitem{c6} Youquan Wang, Lu Zhang , Haicheng Tao , Zhiang Wu, Jie Cao,`A Comparative Study of Shilling Attack Detectors for Recommender Systems,' \emph{IEEE 2015 12th International Conference on Service Systems and Service Management (ICSSSM)},\hskip 1em plus 0.5em minus 0.4em\relax pp. 1-6, June 2015.
\bibitem{c7}
B. Mobasher, R. Burke, R. Bhaumik and J. J. Sandvig, \textit{Attacks and remedies in collaborative recommendation}, IEEE Intelligent Systems, 2007, vol. 22, pp. 56-63.
\bibitem{c8}
H. Xia, B. Fang, M. Gao, H. Ma, Y. Tang, and J. Wen, \textit{A novel item anomaly detection approach against shilling attacks in collaborative recommendation systems using the dynamic time interval segmentation technique}, Information Sciences, 2015, Vol 306, pp.150-165.
\bibitem{c9}
J. Lee, and D. Zhu, \textit{Shilling attack detection—a new approach for a trustworthy recommender system}, INFORMS Journal on Computing, 2011, Vol. 21, pp. 117-131.
\bibitem{c10}
J. Cao, Z. Wu, B. Mao, and Y. Zhang, \textit{Shilling attack detection utilizing semi-supervised learning method for collaborative recommender system}, World Wide Web, 2013,vol. 16, pp. 729-748.
\bibitem{c11}
K. Bryan, M. O'Mahony, and P. Cunningham, \textit{Unsupervised retrieval of attack profiles in collaborative recommender systems}, Technical Report, University College Dublin, 2008.
\bibitem{c12}
C. Li, and Z. Luo, A Hybrid Item-based Recommendation Algorithm against Segment Attack in Collaborative Filtering Systems. 2011 International Conference on Information Management, Innovation Management and Industrial Engineering, pp.403-406, 2011.
\bibitem{c13}
S. Zhang, A. Chakrabarti, J. Ford, and F. Makedon, \textit{Attack detection in time series for recommender systems}, Proceedings of the 12th ACM SIGKDD International Conference on Knowledge Discovery and Data Mining, pp. 809-814, August 2006.
\bibitem{c14}
Chirita PA, Nejdl W, Zamfir C (2005) Preventing shilling attacks in online recommender systems. In: Proceedings of the 7th annual ACM international workshop on Web information
and data management. ACM, pp 67-74
\bibitem{c15}
Guo G, Zhang J, Thalmann D (2012) A simple but effective method to incorporate trusted
neighbors in recommender systems. In: User modeling, adaptation, and personalization. Springer, Heidelberg, pp 114-125
\end{thebibliography}
\end{document}